# CoPhy: A Scalable, Portable, and Interactive Index Advisor for Large Workloads

Debabrata Dash*
ArcSight, an HP Company
ddash@hp.com

Neoklis Polyzotis*
University of California at Santa Cruz
alkis@cs.ucsc.edu

Anastasia Ailamaki
École Polytechnique Fédérale de Lausanne
natassa@epfl.ch

## ABSTRACT

Index tuning, i.e., selecting the indexes appropriate for a workload, is a crucial problem in database system tuning. In this paper, we solve index tuning for large problem instances that are common in practice, e.g., thousands of queries in the workload, thousands of candidate indexes and several hard and soft constraints. Our work is the first to reveal that the index tuning problem has a well structured space of solutions, and this space can be explored efficiently with well known techniques from linear optimization. Experimental results demonstrate that our approach outperforms state-of-the-art commercial and research techniques by a significant margin (up to an order of magnitude).

## 1. INTRODUCTION

The problem of *index tuning* can be loosely defined as follows: Given a workload, a set of candidate indices, and a set of constraints (e.g., a storage budget, a time budget for index construction, or constraints on the characteristics of indexes), select a subset of the candidates that optimize workload evaluation and satisfy the constraints. Index tuning is crucial in database system tuning, since indexes are supported in all major database systems and, if selected carefully, can enable orders of magnitude improvement in workload performance. Naturally, all commercial systems provide automated index tuning methods that assist administrators in this challenging and important task [20, 1].

In this paper, we solve the index tuning problem for large input specifications that are common in real-world systems, e.g., a workload with thousands of statements, thousands of candidate indices, and several soft and hard constraints. Existing index tuning techniques [4, 14, 3] cannot handle such specifications efficiently and hence force administrators to limit the scope of index tuning, e.g., by using a smaller workload, fewer candidate indexes or fewer constraints, or by tuning less frequently. This scale-down increases the work of the administrator and negatively affects the quality of tuning–clearly, a lose-lose situation.

The inefficiency of existing techniques stems from several factors that reflect the challenges behind index tuning. First, the space of possible solutions grows exponentially with the number of candidate indices. This is an inherent property of the problem [7], since indexes interact in terms of their benefit and hence a greedy selection of indexes does not lead to a good solution. Moreover, automated techniques have to rely on *what-if optimization* in order to assess accurately the benefit of an index-set on a specific query. A large input workload, combined with the large solution space, implies a very high number of what-if optimization calls, thus increasing the cost of index tuning dramatically.

Recent methods for fast what-if optimization [15, 5] have been a major development in the area, as they can reduce significantly the cost of index tuning. However, these methods only speed up a single component of the process. The fundamental issue remains that the solution space lacks "structure", i.e., it is necessary to explore an exponential number of index-sets, without much knowledge on the relationship between solutions.

Our work takes a radically different approach to index tuning. The starting point is the aforementioned methods for fast what-if optimization. Instead of treating them like a black box, as is the norm in previous works, we analyze their computation and prove a key theoretical result: *Any index tuning technique that employs fast what-if optimization essentially solves a compact binary integer program (BIP), whose number of variables grows linearly with the size of the input specification.* In effect, we can perform index tuning by employing fast, off-the-shelf BIP solvers, thus taking advantage of more than 60 years of research and development in the field of linear optimization. Ours is the first work to uncover this deep link between index tuning and linear optimization and to show that the solution space is well structured. A previous study [14] considered a similar connection at a much higher level, without exploring the structure of the solution space.

The technical contributions of our work are as follows:

- We prove that index tuning with fast what-if optimization is equivalent to a compact BIP (§3). The derivation of the BIP requires only few what-if optimization calls (to initialize the method for fast what-if optimization). Once derived, the BIP can be solved efficiently with existing methods from linear programming without any what-if optimization calls.
- We develop a novel index tuning technique, termed *CoPhy*, that builds on our BIP-based formulation of the index tuning problem (§4). *CoPhy* is a simple, portable technique that employs an off-the-shelf BIP solver to perform index tuning. Moreover, it leverages well known techniques from linear programming in order to support soft constraints, early termination with quality bounds, and fast re-tuning for small changes to the input specification.
- We demonstrate *CoPhy*'s effectiveness experimentally (§5). *CoPhy* consistently outperforms state-of-the-art commercial and research index tuning techniques [3, 20, 14], both in solution quality and in total execution time.

*The work was done while the first two authors were at EPFL.

## 2. THE INDEX TUNING PROBLEM

We are given a relational database with $n$ tables $T_1, \ldots, T_n$. A *configuration* $X$ is a set of indices defined over the database tables. We do not place any limitations on the indices regarding their type or the type or count of attributes that they cover, except that each index is defined on exactly one table (e.g., no join indices).

We are also given a (representative) workload $W$ that comprises SELECT and UPDATE statements over the database. Given a SELECT statement $q$ in $W$, we use $cost(q, X)$ to denote the cost of the optimal plan to evaluate $q$ assuming that only the indices in $X$ are available. In existing database systems, $cost(q, X)$ can be computed efficiently by invoking the *what-if optimizer*. The latter performs a normal optimization of $q$, "faking" the statistics for the hypothetical indices in $X$. Following common practice, we model an update statement $q$ in $W$ as a query shell $q_r$ that selects the tuples to be updated, and an update shell $q_u$ that performs the update on the selected tuples and also updates any affected indices. We assume that each affected index $a$ has an independent maintenance cost, denoted as $ucost(a, q)$, which can be estimated again through the what-if optimizer. Hence, the total cost of an update statement can be expressed as $cost(q, X) = cost(q_r, X) + \sum_{a \in X} ucost(a, q) + c_q$, where the last term is simply the cost to update the base tuples. In what follows, we use $W_r$ to refer to the SELECT statements and query shells in $W$, and $W_u$ to refer to the update statements in $W$.

Let $S = S_1 \cup \cdots \cup S_n$ be a set of candidate indices, where each $S_i$ contains candidate indices for table $T_i$, $1 \le i \le n$. Set $S_i$ can be derived with automated methods (by analyzing $W$) or manually curated by the administrator. The goal of index tuning is to find the configuration from $S$ that minimizes the total evaluation cost for the representative workload $W$ and also satisfies a set of constraints $C$ (more on that later).

**Index Tuning Problem** *Given a workload $W$ (with a weight $f_q$ per $q \in W$), a set of candidate indices $S$, and a set of constraints $C$, find $X^*$ such that $X^* = \arg\min\{\sum_{q \in W} f_q cost(q, X) \mid X \subseteq S\}$, and $X^*$ satisfies the constraints in $C$.*

The weight $f_q$ can represent the frequency of the statement, or some hand-tuned importance metric specified by the DBA.

The set of constraints $C$ states the required properties for the chosen configuration $X^*$. A common constraint is that the total storage space for $X^*$ does not exceed some budget $M$. A different constraint may be that the total update cost of the selected indices does not exceed some threshold. Yet another constraint may be that $X^*$ should contain at most one clustered index per table. In our work, we allow $C$ to comprise constraints expressed in the rich language proposed by Bruno and Chaudhuri [4]. We also allow for "soft" constraints that may be violated by $X^*$, in the same spirit as [4]. We discuss constraints in more detail in §3.2.

To simplify our notation, we henceforth assume that each statement in $W$ references a specific table $T_i$ at most once[1]. (The extension to the general case is straightforward but the notation gets very complicated.) Under this assumption, a configuration $A \subseteq S$ is called *atomic* [10] if $A$ contains at most one index from each $S_i$. We represent $A$ as an $n$-vector, where the element $A[i]$ is an index from $S_i$ or the special symbol $I_\emptyset$ if no index from $S_i$ is selected. We use $atom(X)$ to denote the set of atomic configurations in $X$.

**Index Advisors.** An *index advisor* is an algorithm that solves an instance of the index tuning problem. At a high level, the index advisor generates a state space of configurations based on $S$, and then invokes a search strategy to find the optimal state. The industrial-strength index advisors typically requires thousands of what-if calls, as it needs to evaluate $cost(q, X)$ for every $q \in W$ and many different choices of $X \subseteq S$.

[1]We consider each tuple variable on $T_i$ as a separate reference.

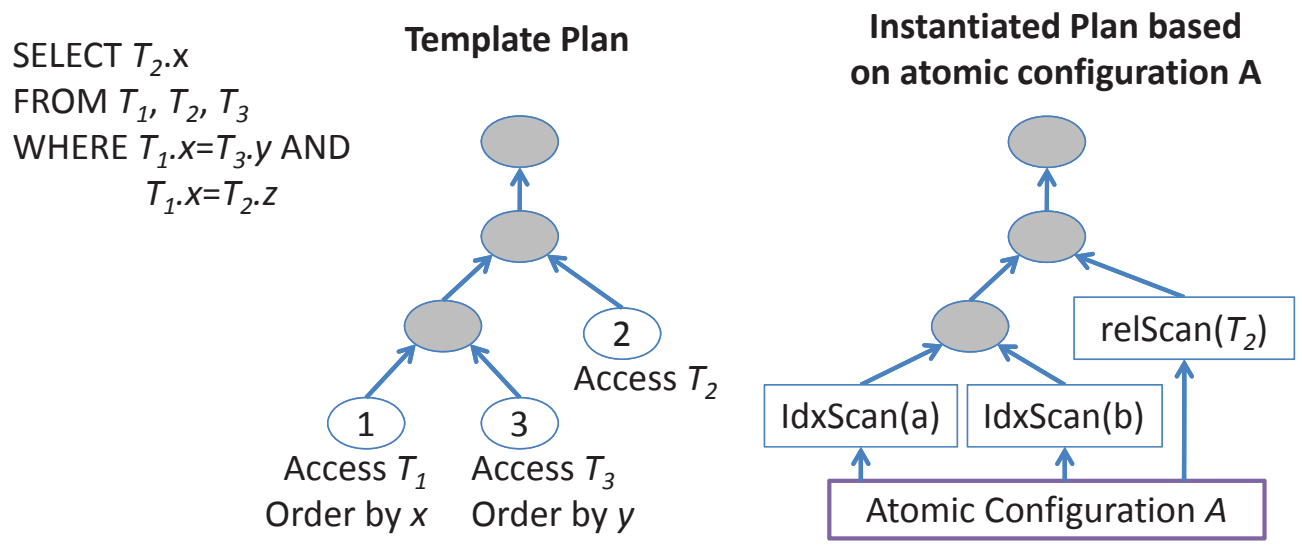

**Figure 1: The left part shows a template plan for a sample join query. The grey ovals denote concrete physical operators, whereas the white ovals denote the holes of the template. Hole $i$ corresponds to the access of table $T_i$ based on a sorted order. On the right part, the template plan is instantiated to a specific physical plan using concrete access methods from an atomic configuration $A$. The configuration $A$ has the following contents: $A[1] = a$, an index with key $T_1.x$; $A[2] = I_\emptyset$; and, $A[3] = b$, an index with composite key $(T_2.x, T_2.w)$.**

Two recent studies have proposed the INUM [15] and C-PQO [5] methods to significantly speedup what-if optimization for the input workload $W$. Both methods implement a what-if optimization interface that wraps the existing what-if optimizer. We describe INUM in more detail because we have more experience at hand, but the same principles apply for C-PQO. Given a workload $W$, INUM preprocesses each SELECT statement and query shell $q$ by making a few carefully selected what-if optimization calls. Using the information gathered from these calls, INUM creates a set of *template plans* for $q$, denoted as $TPlans(q)$, that essentially encode the space of optimal plans for $q$ according to the what-if optimizer. A template plan $p \in TPlans(q)$ is a physical plan for $q$ except that all access methods (i.e., the leaf nodes of the plan) are substituted with "slots". Figure 2 shows an example for a simple query. Given an atomic configuration $A$, there is a unique physical plan that is instantiated from $p$ by "plugging" the slots with access methods from $A$. We use $icost(p, A)$ to denote the cost of this unique plan. Note that $icost(p, A)$ may be infinite if $A$ does not have suitable access methods for all slots in $p$.

Given a configuration $X$, INUM approximates $cost(q, X)$ as the minimum of $icost(p, A)$ for all $p \in TPlans(q)$ and $A \in atom(X)$. As shown in the original study [15], this is an accurate approximation for the purpose of index tuning, and is orders-of-magnitude faster to compute compared to a what-if optimization. Note that searching through the atomic configurations eliminate plans that benefit from index intersections. In the context of index tuing, however, anecdotal evidence suggests that index intersections do not improve the solution quality [5], therefore we ignore their effect in the rest of the paper.

Overall, an index advisor can use INUM or C-PQO to approximate $cost(q, X)$ efficiently and accurately. This approach results in minimal to no loss in the quality of the solution while reducing total execution time of the advisor by several orders of magnitude [5,15]. It is natural to assume that this type of fast what-if optimization will become standard for index advisors, and thus we henceforth consider the index tuning problem when INUM or C-PQO are used to approximate $cost(q, X)$.

## 3. INDEX TUNING ≡ BINARY INTEGER PROGRAM

The main result of our work can be summarized as follows: If $cost(q, X)$ is computed using INUM or C-PQO, then index tun-ing becomes a binary integer program (BIP). The implication is that we can solve the index tuning problem efficiently, robustly, and in a principled fashion, using off-the-shelf, mature BIP solvers. This idea forms the foundation of the *CoPhy* index advisor that we present in the next section.

The result hinges on a specific property of $cost(q, X)$ that we term *linear composability*. We define this property below, and then show that it holds when $cost(q, X)$ is computed by INUM or C-PQO, in the absence of index intersections.

DEFINITION 1. *Function cost is* linearly composable *for a SELECT statement q if there exists an integer $K_q$ and constants $\beta_{qk}$ and $\gamma_{qkia}$ for $k \in [1, K_q], i \in [1, n], a \in S_i \cup I_\emptyset$, such that:*

$$cost(q, X) = \min\{\beta_{qk} + \sum_{i \in [1,n], a = A[i]} \gamma_{qkia}, k \in [1, K_q], A \in atom(X)\},$$

*for any configuration $X$.*

*Function cost is linearly composable for an UPDATE statement q if it is linearly composable for its query shell $q_r$.*

LEMMA 1. *In the absence of index intersections, cost is linearly composable if it is computed by INUM or C-PQO.*

The complete proof appears in the appendix, but we briefly discuss the case of INUM to build some intuition. For INUM, $K_q = |TPlans(q)|$ and each $k$ corresponds to a distinct template plan $p$. In turn, the expression $\beta_{qk} + \sum \gamma_{qkia}$ corresponds to $icost(p, A)$, where $\beta_{qk}$ is the execution cost of the internal operators in $p$ (termed the *internal plan cost* in [15]), and $\gamma_{qkia}$ is the total cost of accessing the table at slot $i$ using the access method $A[i]$.

Linear composability provides a formal characterization for techniques like INUM or C-PQO that pre-process $q$ through the what-if optimizer in order to compute $cost(q, X)$ efficiently for any $X$. The significance of the property is that it casts the selection of the optimal plan for a given $X$ as a minimization over a set of linear expressions. Does this imply that the query optimizer employs a linear cost model? The answer is no. The non-linearity of the optimizer's cost model (which has been shown experimentally [16]) is encoded in the constants $\beta_{qk}$ and $\gamma_{qkia}$ which are query-specific. In other words, linear composability does not limit the application of our techniques in real-world systems.

We develop our main result in the following sub-sections. To facilitate presentation, we first consider the case where $C = \emptyset$, i.e., no constraints are specified, and then extend to the general definition of the index tuning problem.

## 3.1 Base Case: No Constraints

We begin with the formal statement of the main result, and then discuss its implications.

THEOREM 1. *Let $(W, S, C)$ denote an instance of the index tuning problem. If $C = \emptyset$ and cost is linearly composable for every $q \in W$, then solving the index tuning problem is equivalent to solving the following Binary Integer Program (BIP):*

*Minimize:*

$$\sum_{\substack{q \in W_r \\ k \in [1,K_q]}} f_q \beta_{qk} y_{qk} + \sum_{\substack{q \in W_r \\ k \in [1,K_q] \\ i \in [1,n] \\ a \in S_i \cup \{I_\emptyset\}}} f_q \gamma_{qkia} x_{qkia} + \sum_{\substack{q \in W_u \\ a \in S}} f_q z_a ucost(a, q),$$

*For:*

$$x_{qkia} \in \{0, 1\}, y_{qk} \in \{0, 1\}, z_a \in \{0, 1\}$$

*Subject to:*

$$\sum_{k \in [1,K_q]} y_{qk} = 1, \sum_{a \in S_i^+} x_{qkia} = y_{qk}, \; z_a \geq x_{qkia}$$

*The solution to the index tuning problem is computed as $X^* = \{a \,|\, a \in S \wedge z_a = 1\}$.*

The proof is quite lengthy and appears in the appendix (§B). Intuitively, each variable $z_a$ controls the selection of index $a$ in the final configuration $X^*$, while variables $y_{qk}$ and $x_{qkia}$ control the choice for $k$ and $A$ that yield the minimal value for $cost(q, X^*)$ under linear composability (see Definition 1). The constraints ensure that only one $k$ is selected and that $A$ is a member of $atom(X^*)$. Clearly, there is a connection between the linear composability of $cost$ and the final BIP, but, as shown in our proof, the derivation is not obvious.

The theorem has profound implications on the complexity of index tuning in practice.

1 *Any index advisor that employs a linearly composable query cost function (such as the one computed by INUM or CPQ-O) will basically solve a BIP*. Essentially, a linearly composable cost function exposes a significant amount of structure in the index tuning problem, which is absent if the what-if optimizer is treated as a black box. As we discuss in §3.2, we can extend the BIP with additional linear constraints to encode a non-empty $C$, and hence the theorem extends to the general case.

2 *We can use off-the-shelf BIP solvers to perform index tuning*. The advantages are numerous and significant: we leverage the extensive research and development done in the field of linear programming for the past sixty years; we simplify the development of the index advisor; and, we obtain dramatic improvements in scalability and efficiency. Note that the number of variables in the BIP grows linearly with the number of relations $n$, the number of candidate indices in $S$, and the total number of template plans (which typically grows linearly with the size of the workload [15]). Off-the-shelf solvers routinely handle hundreds of thousands of variables, which means that they are very well suited to solve the index tuning problem efficiently.

3 *By relying on fast, mature BIP solvers, we enable an interactive paradigm for index tuning*. A BIP solver can return a solution within minutes (sometimes, within seconds!) even for a large number of variables, thus allowing the administrator to perform several tuning sessions. Moreover, BIP solvers provide continuous feedback on the distance between the currently computed solution and the final solution. The administrator can choose to terminate the tuning session early if the current solution is within an acceptable distance, e.g., within 5% of the final solution, thus reducing further the total time to perform index tuning.

Does the theorem change the complexity of index tuning? The answer is negative. In principle, solving a BIP is NP-Hard and this matches the complexity bound of index tuning [7]. However, off-the-shelf solvers perform much better for real-world problem instances, which is verified in our experimental study.

Our final observation is that the original INUM and C-PQO work focused on the efficiency of what-if optimization, and missed this important connection between the index tuning problem and BIP. The proof of the theorem reuses some of the machinery introduced in [15], but the underlying analysis is one of the novelties of our work.

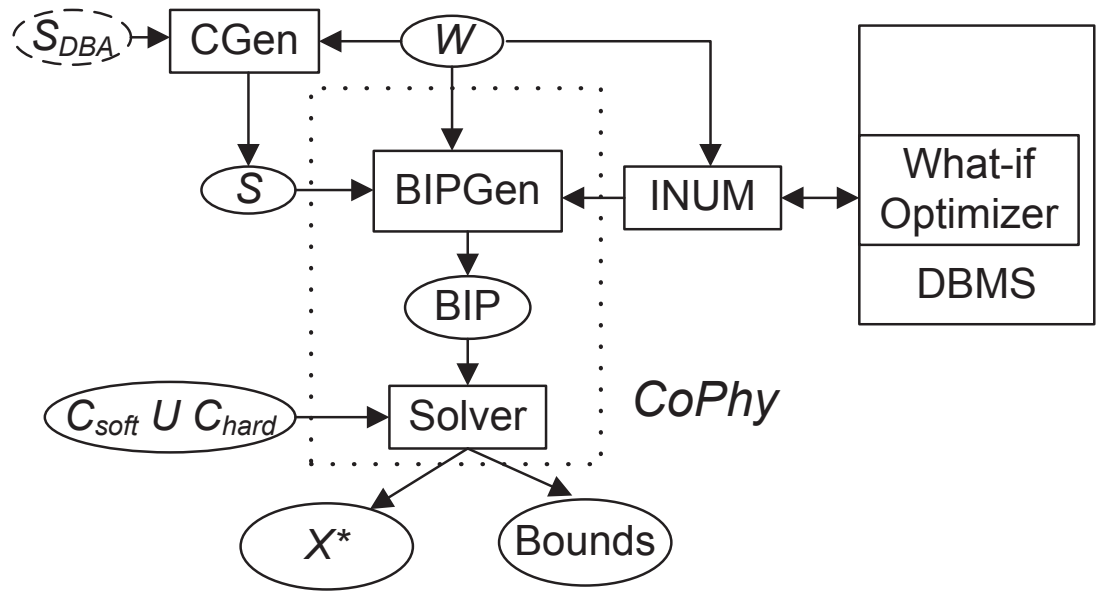

**Figure 2: The architecture of *CoPhy*.**

## 3.2 Adding Constraints

We extend Theorem 1 to the case where $C \neq \emptyset$, i.e., when the optimal configuration must satisfy constraints specified by the database administrator.

Any constraint that can be written in linear form can be incorporated in the BIP of Theorem 1 without further changes. Most practical constraints actually fall in this category, and hence Theorem 1 remains relevant for a large number of real-world scenarios. For instance, in the appendix we discuss how to derive linear constraints for all the use cases on constrained physical design tuning from the study of Bruno and Chaudhuri [4], including but not limited to constraints on individual indexes, on the selected combination, and on the choice of clustered indexes. In fact, every constraint expressed in the language of that study can be converted to a linear constraint that can be added to our BIP, except for non-linear aggregate functions and constraints encoded in UDFs.

To illustrate the gist of our approach, we describe here how to handle the typical constraint of a storage budget. The constraint can be stated as $\sum_{a \in X^*} size(a) \leq M$, where $size(a)$ denotes the estimated size of index $a$, and $M$ is the storage budget. Using the variable $z_a$ that tracks the selection of an index in the final configuration, the storage constraint can be encoded as $\sum_{a \in S} z_a size(a) \leq M$. This new constraint is simply added to the BIP of Theorem 1 without further modifications.

The easy incorporation of additional constraints exemplifies the power of our approach. In particular, it is instructive to make a qualitative comparison to the index advisor developed by Bruno and Chaudhuri [4], which employs a significant amount of machinery on top of an existing index tuning technique in order to handle the same class of constraints. On the other hand, our approach relies on off-the-shelf software to provide similar functionality, and brings features (e.g., feedback on quality of solution, very fast solutions) that are not possible with other techniques.

Currently, our method does not support constraints that cannot be written in a linear form. It may be possible in certain cases to relax the optimization problem in order to incorporate such constraints at the cost of generating an inferior solution, but we have not identified thus far the need for such constraints in practice.

Finally, we note that it is possible to designate some constraints as *soft*, which leads to the generation of several solution along a Pareto-optimal curve. We discuss this issue in the next section, since soft constraints are handled outside of the BIP solver.

# 4. COPHY

This section introduces the *CoPhy* index advisor. *CoPhy*'s foundation is the BIP-based formulation of the index tuning problem that we presented earlier. In addition, *CoPhy* includes a module for candidate-index generation, supports soft constraints, exports an interface for interactive tuning, and allows the DBA to terminate the tuning process early, based on feedback on the quality of the current solution.

Figure 2 shows the high-level design of *CoPhy*. In what follows, we summarize the functionality of each component.

INUM takes $W$ as the input and invokes the DBMS's what-if optimizer to determine $TPlans(q)$ for each query $q$ in $W$. INUM sits outside the DBMS, and uses commonly available APIs to communicate with the DBMS optimizer.

CGen takes as input $W$, examines each query to generate a large number of candidate indices based on the referenced columns, and then forms the candidate set $S$ as the union of the per query sets. The DBA may also specify a set of interesting indices $S_{DBA}$ which are added to $S$. The per query candidates are generated using more or less well known heuristics from the literature. In contrast to existing index advisors, CGen does not apply any complex pruning heuristics, and hence the final set $S$ may be quite large.

BIPGen takes $S$ as input, and the template plans, and builds the BIP using Theorem 1.

Solver takes $C$ and the BIP as inputs, and merges them into another BIP using the methods discussed in §3.2. The expanded BIP is then handed to an off-the-shelf BIP solver to generate the solution $X^*$. Set $C$ contains *hard constraints* ($C_{\text{hard}}$) that must be satisfied by $X^*$, and *soft constraints* ($C_{\text{soft}}$) that conceptually must be satisfied to the extent possible. The Solver also implements the novel features of *CoPhy* mentioned earlier, namely, interactive tuning and early termination.

Overall, *CoPhy* employs off-the-shelf software and generic components and is thus easy to port across several systems. INUM is the only component that interfaces with the DBMS and requires system-specific customization, but it too is easily portable since it relies on two services that are common in modern systems: a what-if optimizer, and the ability to force specific physical plans through hints. Our own implementation experience corroborates these observations: our *CoPhy* prototype is implemented in 2K lines of Java code and supports two very different commercial DBMSs.

The remainder of the section describes the details of the Solver component, which is at the heart of *CoPhy*. We first describe the computation of a solution $X^*$, and then discuss the novel features of early termination and interactive tuning.

**Function** Solver($\mathcal{B}$, $C_{\text{hard}}$)
**Input**: A BIP $\mathcal{B}$ describing the tuning problem instance, and a set of hard constraints $C_{\text{hard}}$.
**Output**: A recommended index set $X^*$
**if** *BIPSolver.isNotFeasible*($\mathcal{B} + C_{hard}$) **then**
  raise InfeasibleException ; `// Problem is infeasible`
$\mathcal{B}_{\mathcal{R}}$ = relax($\mathcal{B}$) ; `// Apply Lagrangian Relaxation`
$X^*$ = *BIPSolver*.solve($\mathcal{B}_R + C_{\text{hard}}$) ; `// Solve the problem`
**return** $X^*$;

**Figure 3: Solver pseudo-code for $C = C_{\text{hard}}$.**

## 4.1 Solver

We describe the main ideas assuming that no soft constraints are specified ($C_{\text{soft}} = \emptyset$) and then discuss the general case.

Algorithm 3 shows the Solver pseudo-code assuming that $C = C_{\text{hard}}$. The Solver takes as input a BIP $\mathcal{B}$, formulated according to Theorem 1, and the set of constraints, and outputs an index set $X^*$ that minimizes $cost(X^*, W)$ and satisfies all the constraints. Internally, the Solver employs an off-the-shelf BIP solver, denoted as *BIPSolver*, which is used to generate the final solution. This internal component is treated as a black box that can be swapped with a better implementation, and hence the Solver (and *CoPhy* in general) can benefit from advances in the area of linear optimization solvers immediately.

The first step in the pseudo-code (line 1) invokes *BIPSolver* to check the feasibility of the input BIP $\mathcal{B}$, i.e., whether any constraints in $C_{\text{hard}}$ cannot be satisfied. This check is supported by all off-the-shelf solvers and is done very efficiently. If it fails, *CoPhy* essentially terminates and reports to the DBA the identified constraints. The DBA has the option of removing the reported constraints or converting them to soft constraints.

The next step (line 3) employs the well known technique of *Lagrangian Relaxation* [11] to transform $\mathcal{B}$ to an equivalent BIP $\mathcal{B}_R$. The purpose of this transformation is to avoid corner cases in $\mathcal{B}$ that would increase the execution time of *BIPSolver*. The details are beyond the scope of the presentation, but the key trick is to move the constraint $\sum_{a \in S_i \cup \{I_\emptyset\}} x_{pia} = y_p$ inside the objective function. Note that this transformation is independent of any other constraints in $C$. Once the relaxed BIP is built, *BIPSolver* is invoked to generate a solution (line 4).

**Handling Soft Constraints.** We describe the incorporation of soft constraints with a simple example. Suppose that the DBA specifies a soft constraint on the storage budget, $\sum_{a \in X^*} size(a) \leq M$. The idea is that a solution $X^*$ can exceed the storage budget $M$ as long it yields a reduced workload cost compared to any other other solution that has a lower total storage. Hence, instead of generating a single solution as in the case of solely hard constraints, the index advisor generates a *a set of solutions* that are *Pareto-optimal* with respect to total workload cost and total index storage. The set of solutions captures the trade-off between storage space and workload cost for storage sizes above $M$.

To generate the Pareto-optimal points efficiently, we take advantage again of the BIP formulation and employ well known techniques from linear optimization [2]. More concretely, we first transform $\mathcal{B}$ in a new BIP $\mathcal{B}'$ that has the objective function $\lambda cost(X, W) + (1 - \lambda)(size(X) - M)$, where $\lambda$ is a parameter with values in $[0, 1]$. Then, we can retrieve all the Pareto-optimal solutions by solving $\mathcal{B}'$ for different values of $\lambda$. (Solving $\mathcal{B}'$ is done using the algorithm in Figure 3.) *CoPhy* employs the Chord algorithm [9] to identify a few values of $\lambda$ that yield a representative subset of the Pareto-optimal set, with provable approximation bounds. This approach quickly generates the solutions that represent best the trade-offs encoded by the soft constraints.

The previous methodology can be trivially extended to several soft constraints. The only difference is the generation of $\mathcal{B}'$, where we introduce several terms in the objective function, one for each soft constraint and with a separate $\lambda$ parameter. Solving $\mathcal{B}'$ and choosing the values for the $\lambda$ parameters is done as previously. The details appear in the appendix.

## 4.2 *CoPhy*'s Novel Features

*CoPhy* supports two novel features, namely early termination and interactive tuning, that improve significantly the usability of the index advisor. These features essentially come for "free" once the BIP is generated, by taking advantage of existing functionality inside the BIP solver.

**Support for Early Termination.** A typical BIP solver first identifies an initial solution that satisfies all the constraints and then gradually improves it in order to minimize the objective function. At any point in time, the solver can report an upper bound on the distance between the current solution and the final solution. *CoPhy* employs this feature to support early termination. Basically, the DBA can examine this bound and decide to terminate the tuning process early if the current solution has acceptable quality. Moreover, the solver can be tuned to return the current solution if it is within a threshold (say, 5%) of the optimal, in order to reduce the overall tuning time without hurting quality.

**Interactive Tuning.** Index tuning is generally an exploratory task, where the DBA may wish to examine recommendations for different choices of $W$, $S$, or $C$. This exploration tends to be incremental, i.e., the DBA applies small changes to the parameters of the tuning problem before re-invoking the index advisor. *CoPhy* takes advantage of the BIP formulation in order to make this incremental exploration very efficient. Specifically, instead of formulating an entirely new BIP, *CoPhy* informs the solver of the delta to the original BIP that correspond to the DBA's changes. The internal solver can then reuse the computation done for solving the original BIP in order to compute an updated solution very efficiently. This decreases significantly the response time of the advisor, thus enabling a highly interactive interface for index tuning.

# 5. EXPERIMENTAL EVALUATION

In this section, we present an experimental study of *CoPhy* using a prototype implementation written in Java v1.6. The implementation comprises 2K LOC, which includes INUM and the *CoPhy*-specific code. We employ CPLEX v12.1 as the external, off-the-shelf BIP solver. The prototype is interfaced with two popular commercial database systems, referred to as System-A and System-B respectively. The two ports, referred to as $CoPhy_A$ and $CoPhy_B$ respectively, have minimal implementation differences, as *CoPhy* relies on standard APIs to connect to the underlying DBMS. Unless otherwise noted, we report on results using $CoPhy_A$.

We conducted several experiments in order to evaluate the performance of *CoPhy* and to answer the following questions:

- How well does *CoPhy* perform compared to state-of-the-art commercial index advisors? (§5.2)
- How well does *CoPhy* perform compared to *ILP*, the state-of-the-art BIP-based index tuning technique? (§5.3)
- What is the effectiveness of *CoPhy*'s novel features, namely, feedback on quality of solution, incremental retuning, and handling of soft constraints. (§5.4)

In what follows, we first describe the experimental methodology and then present the results of the experiments outlined above.

## 5.1 Methodology

**Competitor Techniques.** We compare *CoPhy* against the following techniques.

<u>*ILP* [14]</u> This technique employs an alternative formulation of index tuning as a BIP, and is hence an interesting comparison point for *CoPhy*. At a high level, *ILP* first invokes the what-if optimizer to obtain the cost of each query for a set of candidate atomic configurations, and then constructs a BIP with a distinct variable per candidate configuration. Since the number of such candidate configurations can be large, the researchers introduced pruning techniques to keep the problem size under control [13]. (Contrast this with *CoPhy* which uses a variable per index, thus avoiding pruning entirely.) To ensure a fair comparison, we implement *ILP* in Java using the same solver as *CoPhy*, and we also interface it with INUM so that it can benefit from fast what-if optimizations.

<u>Commercial Index Advisors</u> We compare *CoPhy* against the state-of-the-art index advisors that come with the two commercial database systems. We refer to these advisors as Tool-A and Tool-B respectively. To the best of our knowledge, Tool-A employs the techniques described in [3], and Tool-B employs the techniques in [20]. Also, Tool-B employs a workload compression mechanism in order to cope with large workloads.

| Data Skew ($z$) | Workload | $\frac{perf(X^*_A, W)}{perf(Y^*_A, W)}$ | $\frac{perf(X^*_B, W)}{perf(Y^*_B, W)}$ |
|---|---|---|---|
| 0 | $W^{hom}_{1000}$ | 2.10 | 1.03 |
| 0 | $W^{het}_{1000}$ | 2.29 | 1.64 |
| 2 | $W^{hom}_{1000}$ | 1.37 | 1.02 |
| 2 | $W^{het}_{1000}$ | Tool-A timed out. | 1.58 |

**Table 1: Comparing the effectiveness of *CoPhy* with the commercial advisors. $X^*_s$ is the configuration chosen by *CoPhy* on system $s$. $Y^*_s$ is the configuration chosen by the index advisor.**

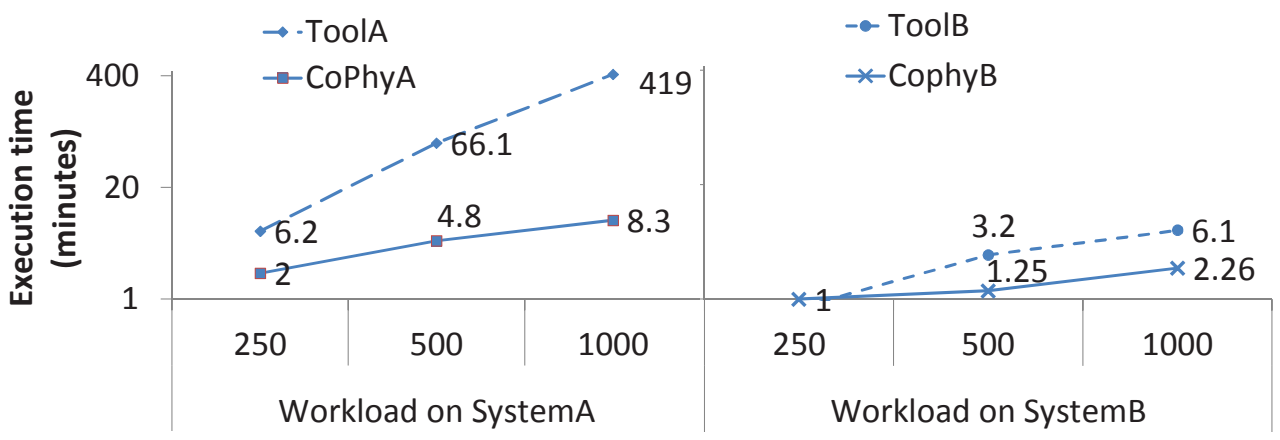

**Figure 4: Comparing the execution times of *CoPhy* with the commercial tools.**

**Data and Queries** We employ a 1GB TPC-H database generated by the *tpcdskew* tool [8]. We generate two versions with skew $z = 0$ (uniform data) and $z = 2$ (highly skewed data).

We study two synthetic workloads with different characteristics. The first workload, denoted as $W^{hom}$, contains random queries generated by the TPC-H query generator on fifteen of the TPC-H query templates. (The remaining seven templates are not supported by the SQL parser of our prototype.) The second workload, denoted as $W^{het}$, is based on a benchmark for index tuning [17] and comprises SPJ queries with group-by and aggregation[2]. Both workloads are sufficiently complex to yield a large set of candidate indices. Workload $W^{het}$ employs many more query templates than $W^{hom}$ and hence represents a heterogeneous workload for which index tuning is more challenging.

We denote a workload of $x$ queries as $W^{hom}_x$ and $W^{het}_x$ for the homogenous and heterogenous workloads respectively, and experiment with $x = 250, 500, 1000$. Intuitively, we expect the cost of an index tuning technique to increase with the workload size. At the same time, the homogeneous workloads have only fifteen distinct query types, and hence favor techniques that employ workload compression, e.g., Tool-B.

**Evaluation Metrics.** We test the various techniques on instances of the index tuning problem that employ the data and workloads described previously and a hard constraint on the total space budget. The latter is expressed as a fraction $M$ of the size of the data.

We measure the effectiveness of a technique through the performance of its recommended configuration $X^*$. Specifically, we report the relative reduction in query processing cost compared to a baseline configuration $X^0$ that contains only the clustered primary key indexes: $perf(X^*, W) = 1 - cost(X^* \cup X_0, W)/cost(X_0, W)$, where $cost(X_0, W)$ and $cost(X^* \cup X_0, W)$ are computed by invoking the what-if optimizer of the corresponding DBMS *directly*. This methodology ensures that the performance of $X^*$ is measured according to the ground-truth of the optimizer's cost model, regardless of any approximations used by the index tuning algorithm. We also report the running time of the technique to generate $X^*$.

**Default Experimental Setup.** Unless otherwise noted, we employ the following defaults for our experiments: $W = W^{hom}_{1000}$, $M = 1$, $z = 0$. The CPLEX solver is tuned to return the first solution that is within 5% of the optimal (we examine this issue in more detail in §5.4). All experiments are executed on a system with a 2.4GHz processor and 2GB of RAM.

## 5.2 Comparison with Commercial Tools

The first set of experiments evaluates *CoPhy* against the commercial index advisors Tool-A and Tool-B. In the interest of space, we summarize a few representative experiments in Table 1 where we vary the complexity of the workload and the data skew. For each system, the table shows the ratio between the improvement $perf(X^*, W)$ for *CoPhy* and the improvement for the index advisor. Hence, a ratio $> 1$ indicates that *CoPhy* yields a better index configuration on the specific system. The detailed results, along with additional experiments for the space budget constraint, are given in the appendix (§C).

Overall, we observe that *CoPhy* consistently outperforms the commercial index advisors, often by a significant margin. Compared to Tool-A, *CoPhy* is better by a margin that ranges from 37% to more than a factor of two. The detailed results indicate that Tool-A has trouble dealing with the high number of queries, whereas *CoPhy*'s BIP-based approach scales gracefully to large workloads.

The comparison to Tool-B shows higher variance and yields interesting insights on the merits of our approach. Specifically, *CoPhy* offers similar performance for the homogeneous workload $W^{hom}_{1000}$, but it becomes significantly better for the heterogeneous workload $W^{het}_{1000}$. Tool-B performs workload compression by sampling, which works well for the homogeneous workload but not so with the more difficult heterogeneous workload. On the other hand, *CoPhy* exhibits stable performance "out of the box" in all the experiments.

For both systems, the gap between *CoPhy* and the commercial tool is reduced when the data is highly skewed ($z = 2$). The reason is that certain indices become very beneficial, which makes it easier to find a good configuration. Still, *CoPhy* continues to perform significantly better compared to the commercial tools.

In terms of the execution time of the three techniques, our detailed experiments reveal that this metric is mostly affected by the size of the workload. Figure 4 shows the execution times as we vary this parameter for $z = 0$ and the homogeneous workload. *CoPhy* is consistently fast for all workload sizes, and the fastest of the three techniques for 500 and 1000 queries. Specifically, *CoPhy* is at least one order of magnitude faster than Tool-A, and is 2x faster than Tool-B for these two workloads. *CoPhy*'s efficiency and stability (w.r.t. the workload variations) is again a result of the BIP-based formulation and the scalability of modern BIP solvers. On the other hand, the commercial index advisors solve the same problem using much less information on the structure of the solution space, which increases significantly their execution times.

**Discussion** The results demonstrate that *CoPhy* operates efficiently and yields index recommendations of high quality. A notable feature of *CoPhy* is that its performance remains stable, whereas Tool-A and Tool-B exhibit high sensitivity to different characteristics of the input workload. Essentially, *CoPhy* alleviates the DBA from the tedious task of selecting a compact representative workload, and eliminates the need for workload compression techniques.

It is also instructive to examine the number of candidate indices examined by the three algorithms. We traced the execution of Tool-A and Tool-B and determined that they used 170 and 45 candidates respectively for the homogeneous workload. These candidate-sets are at least one order of magnitude smaller compared to the 1933 indices examined by *CoPhy*. Again, the power of off-the-shelf BIP solvers allows us to scale-up the index tuning problem. Pruning the candidate-set $S$ becomes less of a concern.

Finally, it is important to stress the simplicity and features of *CoPhy*: it is implemented in 2K LOC, it is portable across two very different systems, it supports soft constraints, and it provides the novel features of early termination and fast retuning (see also §5.4). In contrast, the commercial index advisors are complex software modules, with system-specific code and elaborate search strategies, and do not support the aforementioned features.

[2]We use the C2 query suite with the most complex query templates.

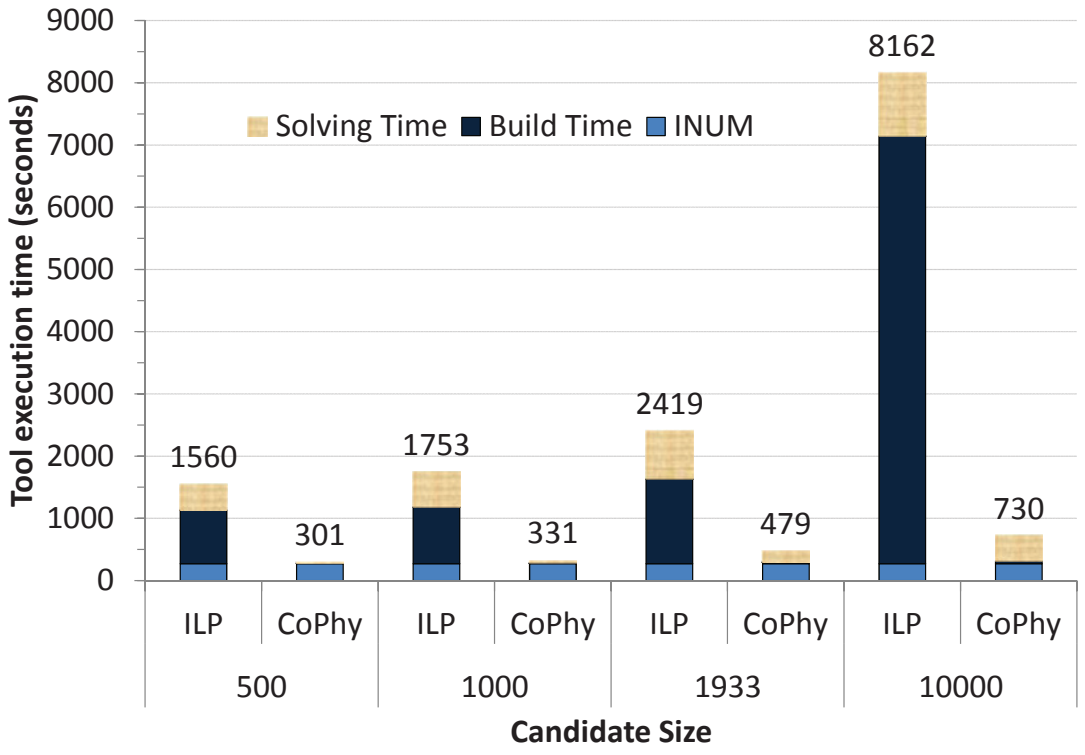

**Figure 5: Execution time of *CoPhy* and *ILP*.**

## 5.3 Comparison with *ILP*

We continue with a comparison between *CoPhy* and *ILP*. As mentioned earlier, *ILP* employs an alternative formulation of index tuning as a BIP. The crucial difference is that *ILP* assigns a variable per atomic configuration, whereas *CoPhy* employs a variable per index. Since the number of atomic configurations grows with $\prod_1^n |S_i|$, *ILP* has to prune substantially the space of atomic configurations before invoking the BIP solver. On the other hand, *CoPhy* builds a much more compact BIP and delegates any pruning to the BIP solver.

Figure 5 shows the performance of *CoPhy* and *ILP* as we vary the size of $S$ (the index candidate set). We consider four possible cases for $S$: $S_{ALL}$ that comprises the 1933 candidates generated by *CoPhy*; two subsets $S_{500}$ and $S_{1000}$ that comprise 500 and 1000 indexes respectively from $S_{ALL}$; and $S_L$ that expands $S_{ALL}$ with random indices for a total of 10K indices. The latter is meant to test the performance of the algorithm for large index sets. We only show total execution time, as the *perf* metric is very similar for the two techniques (actually, *CoPhy* is slightly better by 4-10%).

Clearly, *CoPhy* is dramatically faster compared to *ILP*, with a consistent difference of one order of magnitude. To gain more intuition, we break down the execution time of each technique in three components: *INUM's time* - the time taken to build the INUM's cached plans; *Building time* - the time taken to build the BIP; and finally, *Solving time* - the time taken by the solver to actually find the solutions. As shown, *ILP*'s execution time is dominated by the building time, in which *ILP* performs the pruning of atomic configurations. In comparison, *CoPhy* employs a more principled BIP formulation that does not require any pruning, and which also results in a more compact optimization program. As a result, *CoPhy* spends less time building and solving the BIP, thus scaling up to large sets of candidate indexes.

The experiments for varying workload and storage budget reveal very similar trends, and the results can be found in the appendix.

## 5.4 *CoPhy*-Specific Features

The last set of experiments examine the unique features of the *CoPhy* index advisor. In the interest of space, we report on the most interesting results and defer any details to the appendix.

**Solution Quality Feedback** At any point in time, *CoPhy* can bound the distance between the currently computed solution and the final solution. Figure 6(a) visualizes this bound over time, for three instances of the index tuning problem corresponding to different workloads. As shown, the bound drops fast during the initial iterations over the BIP solver, but then decreases very slowly until the final solution is identified. Using this feedback, the DBA may decide to terminate the tuning session early without compromising significantly the quality of the returned solution. As an example, for $W_{1000}^{hom}$, the DBA can stop the tuning after 4 minutes to obtain a valid solution (satisfying all the constraints) that is at most 5% away from the final solution. Obtaining the final solution requires more than 10 minutes for this workload.

It is interesting to note that *ILP* has a similar feedback capability, since it too relies on a BIP solver. However, the feedback can be generated only after the problem is built, which takes about 30 minutes for $W_{1000}^{hom}$, and even then it will map to the pruned search space of atomic configurations, thus having less intuitive meaning for the DBA.

**Interactive Tuning** We demonstrate next the ability of *CoPhy* to support interactive tuning. We set up the experiment as follows. First, we run *CoPhy* on $W_{1000}^{hom}$ and $S_{1000}$ to obtain an initial recommendation. Subsequently, we augment $S$ with randomly selected indices from $S_{ALL} - S_{1000}$ and request a revised recommendation. This experiment models an interactive scenario where the DBA manually tweaks $S$ in order to fine-tune the recommendation.

Figure 6(b) shows the time to obtain the initial and the revised recommendation for a different number of added indices. The results demonstrate that the solver takes about an order of magnitude less time to solve the BIP for the new candidate indices. Under the hood, the BIP solver is able to reuse some of the initial computation in order to solve the augmented problem. The solving time increases with the number of introduced indices, as expected, but still stays at 55 seconds when 100 new candidates are introduced. Considering the size of the workload, the response time is reasonable in order to support an interactive paradigm. Moreover, a profiling of our prototype showed that *CoPhy* spends about 30 seconds in preprocessing the BIP, an overhead that we may be able to eliminate by re-engineering the BIPGen component.

**Soft Constraints** We examine the effectiveness of *CoPhy* to handle soft constraints. In particular, we substitute the hard constraint on the space budget with a soft constraint $\sum_{a \in X^*} size(a) = 0$, in order to examine the trade-off between the storage budget and the performance of the recommended indices.

Figure 6(c) shows the time to generate five representative points of the Pareto-optimal curve. The points correspond to different values of parameter $\lambda$ (§4.1) that provide a provably good approximation of the overall curve [9]. *CoPhy* has to solve the BIP from scratch for the first point, but it can reuse a significant part of the initial computation to compute the subsequent points. This results in a 4x speed-up compared to a naive re-computation for each point.

The short computation times for subsequent points also enable an interactive exploration of the Pareto-optimal curve based on the DBA's preferences. Returning to the previous experiment, the DBA may select a range of space budgets or a range of quality values where she is interested in exploring the trade-off in more detail. *CoPhy* can compute efficiently the Pareto-optimal points within the selected ranges and thus provide timely feedback to the DBA. This is yet one more example of how *CoPhy* can support an interactive paradigm for index tuning.

# 6. RELATED WORK

Proposed physical design solutions depend heavily on the plan selection mechanism used in the query optimizers. Early research models the query optimizer mathematically [12], and then suggests the design features. Since early optimizers typically use simple cost models [18], it is relatively straightforward to model the entire

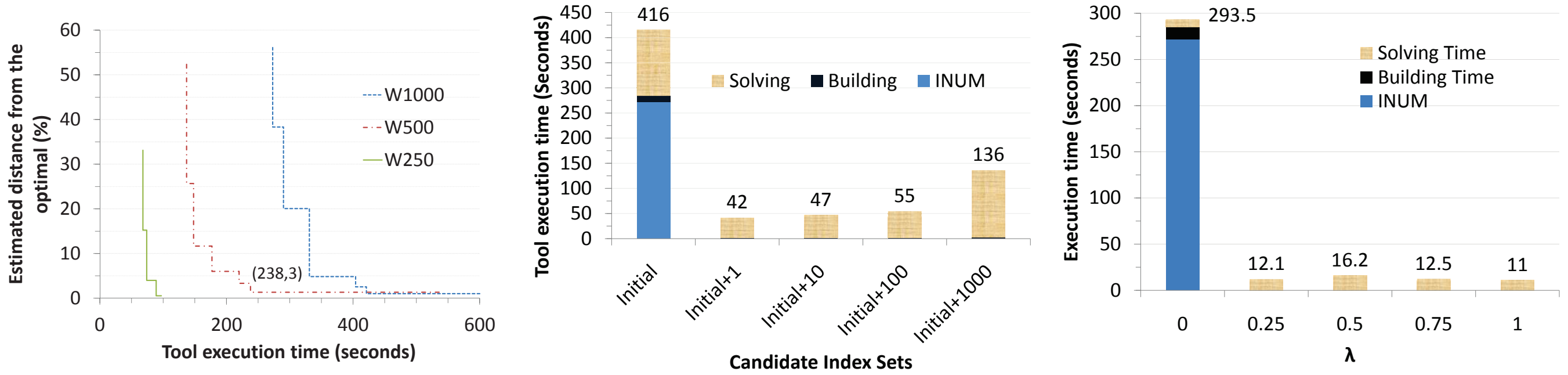

(a) Continuous feedback for early termination. (b) Time to recompute a solution when the candidate-set changes. (c) Time to generate the pareto-optimal curve for a soft constraint.

**Figure 6: *CoPhy*'s interactive features.**

optimization process and select appropriate design features accurately. Modern optimizers, however, use more elaborate cost models which render most previous cost formulations obsolete. Recent work decouples the optimizer design from the problem formulation by modeling the optimizer as the black box and reusing past optimization results [15]. Modeling the optimizer as a black box forces the physical designer to compute the cost of every possible index combination, which is a very expensive process, as such combinations can be exponential in number. We use the same caching approach to model the optimizer, but exploit the internal details of the cache. Our approach allows us to identify useful index combinations without actually enumerating them.

Existing commercial techniques use greedy pruning algorithms to suggest the physical design [20, 1], and use the optimizer directly, thereby reducing their efficiency and predictability. Caprara et al. were the first to propose a BIP approach to the index-selection problem, by modeling it as an extension of the *facility-location problem* (FLP) [6], enabling it to exhaustively search the features, instead of greedily searching them. Their formulation, however, assumes that a query can use only a single index. Papadomanolakis et al. extended the formulation to account for queries using more than one indexes and also model index update costs [14]. Zohreh et al. extend the FLP formulation to use views and then provide heuristics to find optimal physical design in OLAP setting [19]. Their algorithm is tuned towards materializing data-cube views and small number of indexes on them. Our approach scales to an index set two orders of magnitude larger than reported in their work. Since these techniques use FLP formulation, they use heuristic pruning to reduce the problem size to a practical level, a limitation we avoid by proposing a new problem formulation.

# 7. CONCLUSION

In this paper we present *CoPhy*, a practical and scalable index advisor that is based on a novel BIP-based formulation of the index tuning problem. Our experimental results indicate that *CoPhy* outperforms existing techniques by a significant margin and enables a more interactive approach to index tuning. As part of our future work, we would like to extend *CoPhy* to handle other design features such as views and partitions, as well as sequence information for the input workload.

**Acknowledgements** Polyzotis' work was supported in part by the National Science Foundation under Grant No. IIS-1018914.

# APPENDIX

## A. PROOF OF LEMMA 1

We first prove linear composability for INUM. The property follows directly from Equation 1 in [15]. Essentially, the constant $K_q$ is equal to $|TPlans(q)|$ and it represents the size of the template cache in INUM for query $q$. Hence, $k$ corresponds to a specific choice of a template plan $p \in TPlans(q)$. Constant $\beta_{qk}$ denotes the cost of the template's join and aggregation operations (referred to as *internal plan cost*). Each constant $\gamma_{qkia}$ denotes the cost of implementing the access at slot $i$ of the template using some index $a$. For a given index $a \in S_i \cup \{I_\emptyset\}$, the definition of $\gamma_{qkia}$ is as follows:

$$\gamma_{qkia} = \begin{cases} 0, \text{ if } T_i \text{ is not referenced in } q \\ \infty, \text{ if } a \text{ is incompatible with the sorted access on slot } i \\ \text{cost of scan of } T_i, \text{ if } a = \emptyset \\ \text{cost of index scan of } a, \text{ otherwise} \end{cases}$$

Note that $\gamma_{qkia}$ becomes infinite if $a$ is incompatible with the sorted access requirements of slot $i$. Essentially, this means that $a$ is incompatible with the corresponding $p$, therefore encoding the interesting order validity required by INUM. Returning to the example of Figure 2, an index on attribute $T_1.y$ cannot instantiate the access method for slot 1, since the latter requires its output to be sorted on $T_1.x$.

Now, we sketch the proof for linear composability of C-PQO [5]. C-PQO builds a *memo* data structure per query $q \in W$ that encodes all possible plans for different atomic configurations. The memo is essentially an AND-OR graph, where the OR nodes encode the choices that lead to the different optimal plans. Given an atomic configuration $A$, the memo is traversed to identify the OR-choices with minimum cost under $A$.

Given a query $q$, we enumerate the possible OR-choices in the memo and treat each choice as a "template" in INUM's terminology. From that point, the proof is similar as with INUM.

## B. PROOF OF THEOREM 1

We introduce some necessary notation. Given an index configuration $X$, we use $ITcost(X)$ to denote the value of the objective function for the index tuning problem. Similarly, let $\mathbf{v}$ denote a valid (i.e., constraint-satisfying) assignment to the variables of the BIP. We use $BIPcost(\mathbf{v})$ to denote the value of the objective function under the specific assignment. We also use $\mathbf{v}(y_{qk})$ to denote the value of the variable in the assignment, and define $\mathbf{v}(x_{qkia})$ and $\mathbf{v}(z_a)$ similarly.

We observe that the objective function of the index tuning problem has a fixed component $\sum_{q\in W_u} c_q$ that represents the cost to update the base tuples and it does not depend on the choice of $X$. Hence, we ignore this component in what follows.

The proof of the theorem works in two steps. First, we show that every configuration $X$ is mapped to an assignment $\mathbf{v}_X$ such that $ITcost(X) = BIPcost(\mathbf{v}_X)$. This property guarantees that the solution space of the BIP contains all possible solutions for the index tuning problem. Subsequently, we prove that the optimal assignment $\mathbf{v}$ can be mapped to the configuration $X^* = \{a \,|\, a \in S \wedge \mathbf{v}^*(z_a) = 1\}$ such that $ITcost(X^*) = BIPcost(\mathbf{v}^*)$. Combined with the inclusion property of the search space, this property guarantees that $X^*$ is indeed the optimal solution to the index tuning problem. This concludes the proof of the theorem.

We prove the aforementioned properties in the following two lemmata. In both proofs, we make use of the fact that *cost* is linearly composable for every $q \in W$.

LEMMA 2. *For any $X \subseteq S$, there is an assignment $\mathbf{v}_X$ such that such that $ITcost(X) = BIPcost(\mathbf{v}_X)$.*

PROOF. We can rewrite $ITcost(X)$ as follows:

$$ITcost(X) = \sum_{q\in W_r} f_q\, cost(q, X) + \sum_{\substack{q\in W_u\\ a\in X}} f_q\, ucost(a, q)$$

For any $q \in W_r$, linear composability guarantees that $cost(q, X) = \beta_{qk} + \sum_{i\in[1,n],a=Y[i]} \gamma_{qkia}$ for some choice of $k = k_q \in [1, K_q]$ and $Y = Y_q \in atom(X)$. We define assignment $\mathbf{v}_X$ as follows: $\mathbf{v}_X(y_{qk}) = 1$ if $k = k_q$; $\mathbf{v}_X(x_{qkia}) = 1$ if $k = k_q$ and $a = Y_q[i]$; $\mathbf{v}_X(z_a) = 1$ if $a \in X$. In all other cases, the corresponding variable is set to 0. It is straightforward to verify that $\mathbf{v}_X$ satisfies the constraints of the BIP. Moreover, if we expand the BIP objective function for this specific assignment, we get the following sequence of equalities:

$$\begin{aligned} &BIPcost(\mathbf{v}_X) \\ &= \sum_{\substack{q\in W_r\\ k=k_q}} f_q\beta_{qk} + \sum_{\substack{q\in W_r\\ k=k_q\\ i\in[1,n]\\ a=Y_q[i]}} f_q\gamma_{qkia} + \sum_{\substack{q\in W_u\\ a\in X}} f_q\, ucost(a, q) \\ &= \sum_{\substack{q\in W_r\\ k=k_q}} f_q \left( \beta_{qk} + \sum_{\substack{i\in[1,n]\\ a=Y_q[i]}} \gamma_{qkia} \right) + \sum_{\substack{q\in W_u\\ a\in X}} f_q\, ucost(a, q) \end{aligned}$$

The expanded expression for $ITcost(X)$ that we wrote above is equal to the last expression, which completes the proof of the lemma. □

LEMMA 3. *Let $\mathbf{v}^*$ denote the solution to the BIP of Theorem 1. Then, $ITcost(X^*) = BIPcost(\mathbf{v}^*)$ for $X^* = \{a \,|\, a \in S \wedge \mathbf{v}^*(z_a) = 1\}$.*

PROOF. The assignment must satisfy the constraints of the BIP, which implies that $\mathbf{v}^*(y_{qk}) = 1$ for exactly one choice of $k$ per $q$. Let $k^*_q$ denote this choice, and let $Y^*_q$ denote the atomic configuration that corresponds to $\mathbf{v}^*(x_{qkia}) = 1$ for $k = k^*_q$. Then, by eliminating the variables with value 0 and by applying standard algebraic manipulations, we can express $BIPcost(\mathbf{v}^*)$ as follows:

$$BIPcost(\mathbf{v}^*) = \sum_{\substack{q\in W_r\\ k=k^*_q}} f_q \left( \beta_{qk} + \sum_{\substack{i\in[1,n]\\ a=Y^*_q[i]}} \gamma_{qkia} \right) + \sum_{\substack{q\in W_u\\ a\in X^*}} f_q\, ucost(a, q)$$

First, we show that the lemma holds if $k^*_q$ and $Y^*_q$ correspond to the choices of $k$ and $Y$ respectively that minimize $cost(q, X^*)$ under linear composability (Definition 1). Under this assumption, the sum within parentheses in the previous expression is precisely $cost(q, X^*)$. The final equality to $ITcost(X^*)$ is trivial to show.

To conclude the proof of the lemma, we show that this is the only case for $cost(q, X^*)$, i.e., its minimum is achieved for $k = k^*_q$ and $Y = Y^*_q$. We prove this claim by contradiction. Suppose that there exists a different choice $k = \kappa_q \in [1, K_q]$ and $Y = \Psi_q \in atom(X^*)$ such that:

$$\beta_{q\kappa} + \sum_{i\in[1,n],a=\Psi_q[i]} \gamma_{q\kappa ia} < \beta_{qk^*_q} + \sum_{i\in[1,n],a=Y^*_q[i]} \gamma_{qk^*_q ia}. \quad (1)$$

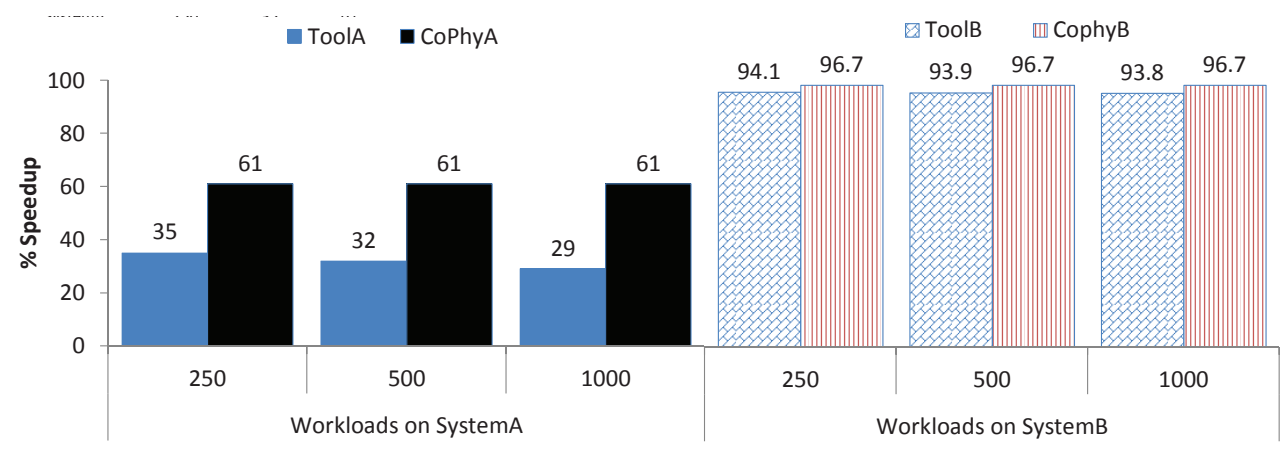

**Figure 7: Quality of solution.**

We define the following assignment $\mathbf{v}'$: $\mathbf{v}'(z_a) = \mathbf{v}^*(z_a)$; $\mathbf{v}'(y_{qk}) = 1$ if $k = \kappa_q$; $\mathbf{v}'(x_{qkia}) = 1$ if $k = \kappa_q$ and $a = \Psi_q[i]$. In all other cases, the corresponding variable is 0. We can verify that this assignment satisfies the constraints of the BIP and is therefore a valid solution. By rewriting $BIPcost(\mathbf{v}')$ to eliminate variables of value 0, and then applying the inequality in Equation 1 for each $q \in W_r$, we have the following sequence of inequalities:

$$BIPcost(\mathbf{v}') =$$

$$\sum_{\substack{q \in W_r \\ k = \kappa_q}} f_q \left( \beta_{q_k} + \sum_{\substack{i \in [1,n] \\ a = \Psi_q[i]}} \gamma_{qkia} \right) + \sum_{\substack{q \in W_u \\ a \in X^*}} f_q ucost(q, a) <$$

$$\sum_{\substack{q \in W_r \\ k = k_q^*}} f_q \left( \beta_{q_k} + \sum_{\substack{i \in [1,n] \\ a = Y_q^*[i]}} \gamma_{qkia} \right) + \sum_{\substack{q \in W_u \\ a \in X^*}} f_q ucost(q, a)$$

$$= BIPcost(\mathbf{v}^*)$$

This contradicts our assumption that $\mathbf{v}^*$ forms an optimal assignment. □

# C. ADDITIONAL EXPERIMENTS

This section provides more details on experimental results described in §5 comparison between *CoPhy* and the commercial tools, then comparison with *ILP*.

## C.1 Comparison with Commercial Tools

**Effect of Workload Size** Figure 7 shows the solution quality for each algorithm for the three workloads $W_{250}^{hom}$, $W_{500}^{hom}$ and $W_{1000}^{hom}$. The results show that *CoPhy* generates consistently solutions of high quality and its performance is basically immune to the size of the workload. Tool-B exhibits similar trends, whereas Tool-A's performance degrades very quickly for large workloads. *CoPhy* also generates the recommendations of highest quality across all techniques, outperforming Tool-A by an order of magnitude and Tool-B by a small margin.

**Effect of Space Budget** In this experiment we vary the space budget from 0.5 to 2 and observe the ratio of the solution qualities for the tools and *CoPhy* on $W_{1000}^{hom}$. Figure 8 shows that, *CoPhy* again provides better quality solutions compared to Tool-A and Tool-B for all space budgets.

**Effect of Workload Diversity.** Figure 9 shows the performance of *CoPhy* and Tool-B for workloads $W_{250}^{het}$, $W_{500}^{het}$ and $W_{1000}^{het}$. We observe that *CoPhy* outperforms Tool-B by a significant margin in all cases. In this case, Tool-B does not benefit as much from its workload compression technique, which is based on random sampling. Hence, its performance drops significantly compared to Figure 7.

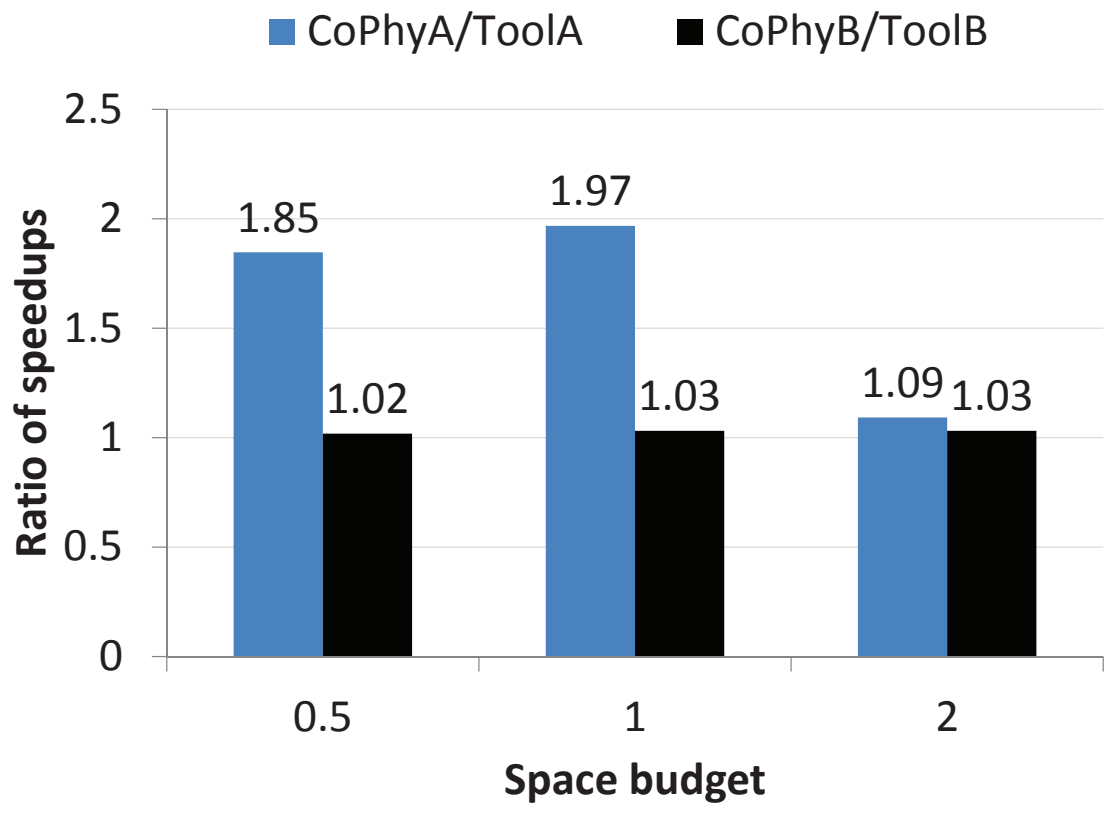

**Figure 8: Speedups for various space budgets.**

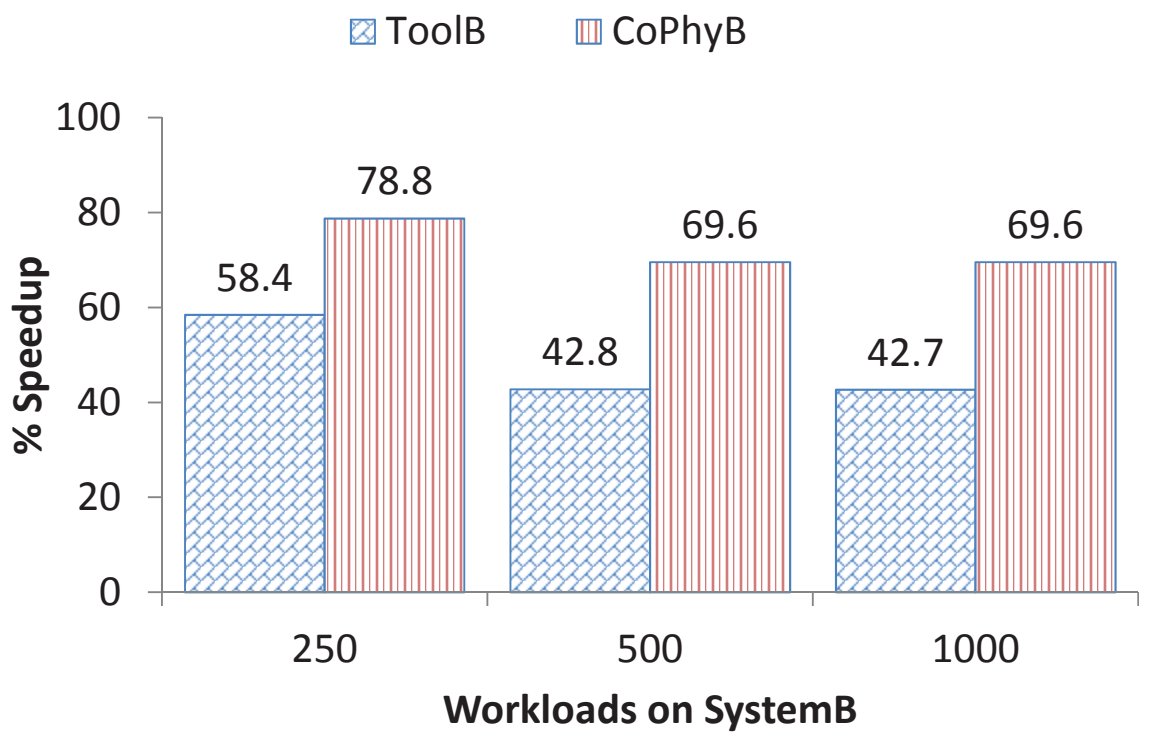

**Figure 9: Speedups on System-B with a diverse workload.**

*CoPhy* also exhibits a drop in performance compared to the homogeneous workload, but overall it generates solutions of high quality.

**Effect of Data Skew.** We have already reported on the experiments with $z = 0$ and $z = 2$. For $z = 1$ and $W = W_{1000}^{hom}$, Tool-Asuggests indexes that provide 67% speedup, compared to 92% for the indexes suggested by $CoPhy_A$. We observe a similar trend in Tool-B: its indices provide 96.9% speedup while $CoPhy_B$ provides 98.1%.

## C.2 Comparison with *ILP*

Figure 10 shows that *CoPhy* outperforms *ILP* by at least a factor of 5 for all workload sizes. Ignoring the INUM cache population time (common to both the techniques), *CoPhy* is typically an order of magnitude faster than *ILP*. *ILP*'s execution time is dominated by the process of pruning the atomic configurations, and *CoPhy* scales by letting the solver systematically search and prune the atomic configurations.

# D. BUILDING THE PARETO-CURVE

We use the Chord algorithm [9], to incrementally build the skyline, while minimizing the number of solver invocations. The algorithm builds an approximation of the skyline as a set of plane segments. The main intuition behind the algorithm is: *given two* $\vec{\lambda}$ *values,* $\vec{\lambda_a}$ *and* $\vec{\lambda_b}$*, and their corresponding objective values* $\vec{f_a}$*, and* $\vec{f_b}$*; the distance of the plane defined by* $(\vec{f_a}, \vec{f_b})$ *and the skyline is maximum at* $\lambda_s = slope(\vec{f_a}, \vec{f_b})$.

The algorithm starts with invoking the solver $j = 0, \ldots, m$ times by setting $\lambda_j = 1, \lambda_k = 0, if k \neq j$. It then finds the pareto-optimal point $\vec{f_s}$ at $\lambda_s$, and computes the distance of the point from the plane defined by the initial invocation. If the distance is less than an acceptable value $\epsilon$, then the algorithm terminates. If not, the algorithm recurses on the plane segments $(\vec{f_a}, \vec{f_s})$ and $(\vec{f_s}, \vec{f_b})$. The set of plane segments returned by the algorithm is always at most $\epsilon$ way from the skyline. Since each iteration of the algorithm changes only the objective, the solver reuses computations from the first invocation to speed up the subsequent optimizations. Thus, the skyline is built interactively and efficiently.

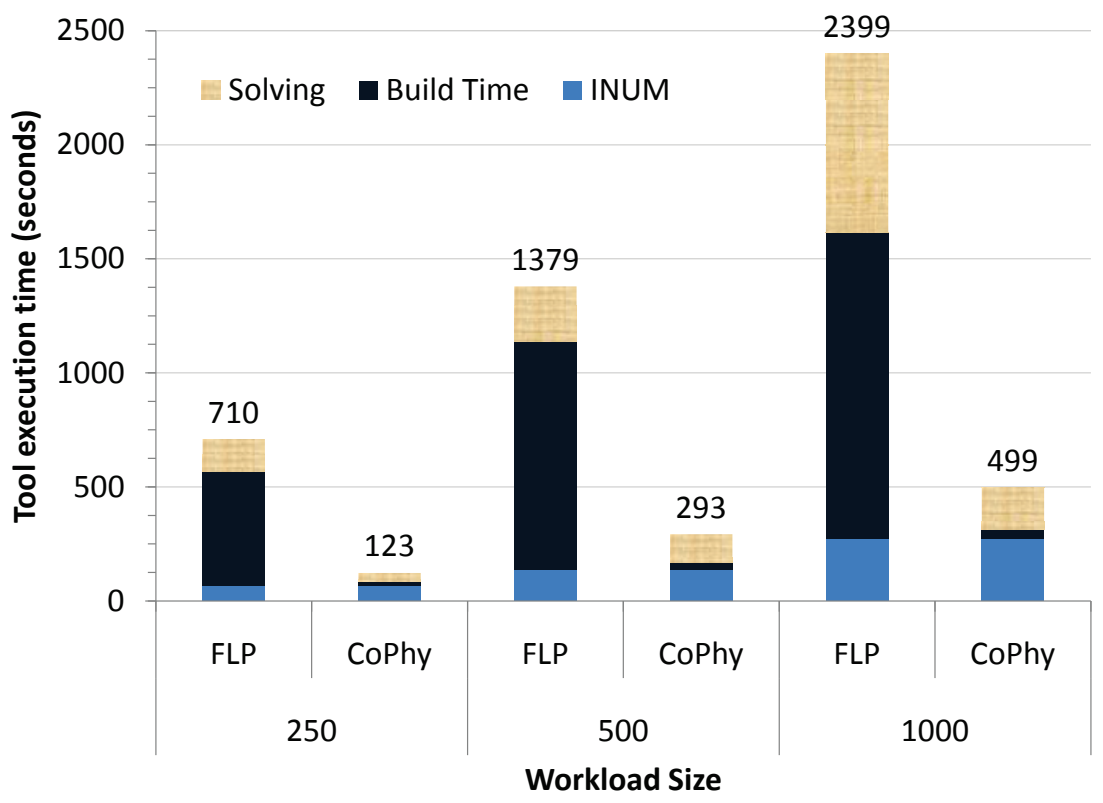

**Figure 10: Comparison of execution times for various workload sizes.**

# E. MORE REAL-WORLD CONSTRAINTS

Section 3.2 discussed incorporating the storage constraint in the BIP. In this section, we extend the BIP to incorporate all real-world constraints reported by Bruno et al. [4]. We group the reported constraints into four categories: index constraints, query cost constraints, generators, and describe their translations into the BIP. We also discuss the scenario when the constraint cannot be translated to the BIP.

## E.1 Index Constraints

This type of constraints specify conditions on the indexes in $X^*$, such as conditions on their size, contained columns, or their column-width. If the DBA constraint applies to a subset $S_c \subset S$ of the candidate indexes, and each index is associated with $w_a$ coefficient, the following linear BIP constraint can be used:

$$\sum_{a \in S_c} z_a w_a <=> V \quad (2)$$

The operator $<=>$ represents the comparison operator the DBA wants to enforce on the indexes w.r.t. the constant $V$. For the size constraint discussed in Section 3.2, $S_c = S$, $w_a = size(a)$, and $V = M$ with $\leq$ as the comparison operator. The constraint is linear, therefore can be easily accommodated into the BIP.

In another instance, the DBA can specify that at most 2 indexes containing more than 5 columns should be selected on the table $T_i$. Here the set $S_c$ contains the indexes on $T_i$ with more than 5 columns, $w_a = 1$, and $V = 2$. Using these assignments, the following constraint is added to the the BIP:

$$\sum_{a \in S_c} z_a < 2 \quad (3)$$

## E.2 Query Cost Constraints

The DBA can specify any constraint on the query costs. For example, she may want to make sure that $X^*$ speeds up all queries in the workload by at least 25% compared to the initial index set $X_0$. We translate such DBA constraints to BIP constraints of the following form:

$$cost(q, X^*) \leq 0.75\ cost(q, X_0) \quad (4)$$

Since the cost function is linear, the constraint also remain linear.

## E.3 Generators

Generators allow the DBA to specify constraints for each query, index, or table, without specifically mentioning them. It is equivalent to *for-loops* in regular programming languages. For example, if the DBA wants to restrict the final query cost, she specifies the following constraint:

FOR $q$ IN $W$
ASSERT $cost(q, X^*) \leq 0.75\ cost(q, X_0)$

The syntax of the constraint is self-explanatory. This DBA constraint is translated by adding constraints shown in Eq. 4 for each query in $W$. The generator can optionally contain *Filters* to limit the scope of the constraints.

For instance, an important implicit constraint on the index tuning problem is: every table in the database can support at most one clustered index. Here the filter prunes out the non-clustered indexes, and the generator iterates over all tables in the database. If the $clustered(S_i)$ represents the set of clustered indexes on $T_i$, then the clustered index constraint can be translated to the following linear BIP constraint:

$$1 \leq i \leq n, \sum_{a \in clustered(S_i)} z_a \leq 1 \quad (5)$$

## E.4 Aggregation and Nested Constraints

Aggregation operations such as SUM, COUNT, MIN, MAX can be accommodated using techniques similar to the index size constraint in §3.2. Similarly, nested constraints are straightforward extensions of the generators. They can be translated into the BIP by determining each instance of the constraint and then converting it into the corresponding BIP constraint (very similar to the loop unrolling technique used in compilers).

## E.5 Non-linear Constraints

Though not seen in practice, the DBA may specify constraints which are non-linear. Such constraints can be in the form of complex user defined functions (UDF), therefore are black-boxes that need to be probed for various index combinations. There are two approaches to address these constraints: First, if the UDF constraint can be approximated by a set of linear constraints, then the approximation can be directly plugged into the BIP. Second, the BIP-Solver's search algorithm can be augmented to reject optimal solutions unless they satisfy the UDF constraint. The latter approach will lose some efficiency compared to a pure linear BIP, but will be faster than the ad-hoc optimization algorithms as it restricts the space of the solutions using non-UDF constraints.

**Summary:** This section demonstrates the generality of the BIP formulation, by translating numerous complex constraints. All reported real-world constraints are linear in nature, allowing efficient solutions to the BIP. In presence of complex black-box constraints the BIP takes advantage of the linear constraints, and exhaustively searches the black-box constraint to suggest valid solutions.